\documentstyle[12pt]{article}
\begin{document}

\title{Schr\"{o}dinger's Interpolating Dynamics and
Burgers' Flows}

\author{Piotr Garbaczewski\thanks{Presented by P. G. at the
International Conference on  Applied Chaotic Systems, Inow{\l}odz,
Poland, September 26-30, 1996}, Grzegorz Kondrat and Robert Olkiewicz\\
Institute of Theoretical Physics, University of Wroc{\l}aw,
Pl. M. Borna 9,\\
PL-50 204 Wroc{\l}aw, Poland}

\maketitle

\begin{abstract}
 We discuss  a connection (and a proper place in this framework)
 of the unforced and deterministically
forced Burgers equation for local velocity fields of certain flows,
with probabilistic solutions
of the so-called  Schr\"{o}dinger interpolation problem. The latter
allows to reconstruct the microscopic dynamics of the system from the
available probability density data, or the input-output statistics
in the phenomenological situations.
An issue   of deducing the most likely
 dynamics (and matter transport) scenario from the given initial
 and terminal probability density data, appropriate e.g. for
studying chaos in terms of densities,  is  here exemplified
 in conjunction with Born's statistical interpretation
postulate in quantum theory, that  yields stochastic processes
which are compatible with the  Schr\"{o}dinger picture
free quantum evolution.
\end{abstract}

\section{The Schr\"{o}dinger reconstruction problem:
most likely microscopic dynamics from the input-output statistics data}

Probability measures, both invariant and nontrivially time-dependent,
often  on different levels of abstraction, are
ubiquituous in diverse areas of physics.
According to pedestrian intuitions, \cite{kac}, one normally expects
that any kind of time developement (dynamics, be it
deterministic or random), which is analyzable in terms
of probability, under suitable mathematical restrictions may give rise to
a well defined  stochastic process. Non-Markovian implementations are
regarded to be  close to reality, but  the corresponding Markovian
approximations  (when appropriate) are easier to handle analytically.

Given a dynamical law of motion (for a particle as example), in many
cases one can associate with it (compute or approximate the observed
frequency data) a probability distribution and various mean values. In
fact, it is well known that inequivalent finite difference random motion
problems may give rise to the same continuous approximant (like e.g.
in case of the
diffusion  equation representation of discrete processes).
As well, in the
study of nonlinear dynamical systems, specifically those exhibiting
the so-called deterministic chaos, \cite{glass,mackey,beck},
given almost any (basically one-dimensional in the cited references)
probability density, it is possible to construct an infinite number
of deterministic finite difference equations, whose iterates are
chaotic and which give rise to this a priori prescribed density. \\
Studying dynamics in terms of densities of probability measures
instead of individual paths (trajectories) of a physical system
is a respectable tool,
\cite{mackey}, even if we know exactly the pertinent microscopic
dynamics.\\
Under general circumstances, the main task of a physicist is
 to fit a concrete dynamical model (through a clever guess or else)
 to  available  phenomenological data.
Then, the distinction between the chaotic (nonlinear, deterministic)
and purely stochastic implementations may not be sharp enough to allow for
a clean discrimination between those options:
the intrinsic  interplay between the stochastic and deterministic
modelling of physical phenomena, \cite{beck},  blurs the access to reality
and certainly precludes a definitive choice of one type of modelling
against the other.

An \it inverse \rm operation of deducing the detailed (possibly
individual, microscopic) dynamics, which is either
compatible with a given  probability measure (we shall be mostly
interested in those admitting densities) or  induces
its own time evolution, cannot  have a unique solution.
However, the level of
ambiguities can be substantially reduced, if we invoke the so-called
 Schr\"{o}dinger problem of reconstructing the  microscopic dynamics
from the given input-output statistics data and/or from the a priori known
time developement  of a given  probability density. The problem   is
known to give rise to
a particular class of solutions (\it most likely \rm interpolations),
in terms of Markov diffusion processes, \cite{mis}-\cite{olk1}. \\
In its  original formulation, due to Schr\"{o}dinger,
\cite{schr}-\cite{zambr},
one seeks the answer to the following question: \\
 given two strictly positive (usually on an open space-interval)
 boundary probability densities  $\rho _0(\vec{x}), \rho _T(\vec{x})$ for
 a process with the time of duration $T\geq 0$.
 Can we \it uniquely \rm identify the stochastic process
interpolating between them ?

Another version of the same problem, \cite{mis}, departs from a given
(Fokker-Planck-type)  probability density evolution and investigates
 the circumstances allowing to deduce a unique random process from
 this dynamics.  We shall pay some attention to this issue in Section 3.

The answer to the above Schr\"{o}dinger's quaestion is known to
be affirmative, if we assume the interpolating
process to be Markovian.
In particular, we can get here a unique Markovian
diffusion process which is specified by the joint probability distribution
$${m_T(A,B) = \int_A d^3x\int_B d^3y\,  m_T(\vec{x},\vec{y})}\eqno{(1)}$$
$$\int d^3y\,  m_T(\vec{x},\vec{y}) = \rho _0(\vec{x}) $$
$$ \int d^3x\,  m_T(\vec{x},\vec{y})=\rho _T(y)$$
where
$${m_T(\vec{x},\vec{y}) = u_0(\vec{x})\, k(x,0,y,T)\, v_T(\vec{y})}
\eqno{(2)}$$
and the two unknown  functions
$u_0(\vec{x}), \, v_T(\vec{y})$ come out as solutions of \it the same sign
\rm of the integral identities (1). Provided, we have at our disposal a
continuous bounded  strictly positive (ways to relax this assumption were
discusssed in Ref. \cite{acta}) integral kernel
$k(\vec{x},s,\vec{y},t),0\leq s<t\leq T$.

We shall confine further
attention to cases governed by the familiar Feynman-Kac kernels.
Then,  the solution of the Schr\"{o}dinger boundary-data problem
in terms of the interpolating  Markovian diffusion process is found
to be completely specified by the   adjoint pair  of parabolic equations.
In case of gradient forward drift fields,  the pertinent
process can be determined
by   checking  (this imposes limitations on the admissible 
potential) whether  the Feynman-Kac kernel
$${k(\vec{y},s,\vec{x},t)=\int exp[-\int_s^tc(
\vec{\omega }(\tau ),\tau)d\tau ]
d\mu ^{(\vec{y},s)}_{(\vec{x},t)}(\omega )}\eqno (3) $$
is positive and continuous in the open space-time
area of interest (then, additional limitations on the path measure
need to be introduced, \cite{blanch}), and whether it
gives rise to
positive solutions of the adjoint pair of  generalised 
heat equations:
$${\partial _tu(\vec{x},t)=\nu \triangle u(\vec{x},t) -
c(\vec{x},t)u(\vec{x},t)}\eqno (4)$$
$$\partial _tv(\vec{x},t)= -\nu \triangle v(\vec{x},t) +
c(\vec{x},t)v(\vec{x},t)\enspace .$$
Here, a function $c(\vec{x},t)$
is restricted only by the positivity and continuity demand
for the kernel (3),  see e.g.  \cite{olk2}.
In the above, $d\mu ^{(\vec{y},s)}_{(\vec{x},t)}(\omega)$ is the
conditional
Wiener measure over sample paths of the standard Brownian motion.

Solutions of (4), upon suitable normalisation give rise to the
Markovian  diffusion process with the \it factorised \rm
probability density $\rho (\vec{x},t)=u(\vec{x},t)v(\vec{x},t)$
which, while evolving in time, interpolates between
the boundary density data $\rho (\vec{x},0)$ and 
$\rho (\vec{x},T)$. The interpolation admits a realisation in terms
of Markovian diffusion processes  with the respective forward and
backward drifts  defined as follows:
$${\vec{b}(\vec{x},t)=2\nu {{\nabla v(\vec{x},t)}
\over {v(\vec{x},t)}}}
\eqno (5)$$
$$\vec{b}_*(\vec{x},t)= - 2\nu {{\nabla u(\vec{x},t)}
\over {u(\vec{x},t)}}$$
in the prescribed time interval $[0,T]$.\\
The related  transport equations for the densities easily follow.
For the forward interpolation, the familiar Fokker-Planck
equation holds true:
$${\partial _t\rho (\vec{x},t) = \nu \triangle
\rho (\vec{x},t) - \nabla [\vec{b}(\vec{x},t)\rho (\vec{x},t)]}
\eqno (6)$$
while for the backward interpolation we have:
$${\partial _t\rho(\vec{x},t) = - \nu \triangle \rho (\vec{x},t) -
\nabla [\vec{b}_*(\vec{x},t) \rho (\vec{x},t)]\enspace . }\eqno (7)$$

We have assumed that drifts are
gradient fields, $curl \, \vec{b}= 0$. As a consequence,
those  that are allowed by the prescribed choice of
$c(\vec{x},t)$  \it must \rm fulfill the compatibility
condition
$${c(\vec{x},t) = \partial _t \Phi \, +\,
{1\over 2} ({b^2
\over {2\nu }}+ \nabla b)}\eqno (8)$$
which establishes the Girsanov-type  connection of
the forward drift
$\vec{b}(\vec{x},t)=2\nu \nabla \Phi (\vec{x},t)$ with the
Feynman-Kac,
cf. \cite{blanch,olk2}, potential $c(\vec{x},t)$.
In the considered Schr\"{o}dinger's interpolation
framework, the forward and backward
drift fields  are  connected   by the identity
$\vec{b}_*= \vec{b} - 2\nu \nabla ln \rho $.

One of the distinctive features of Markovian diffusion processes
with the positive density $\rho (\vec{x},t)$ is that, given the
transition probability density of the (forward) process,  the
notion of the \it backward \rm transition probability density
 $p_*(\vec{y},s,\vec{x},t)$ can be consistently introduced on 
each finite  time interval, say $0\leq s<t\leq T$:
 $${\rho (\vec{x},t) p_*(\vec{y},s,\vec{x},t)=p(\vec{y},s,\vec{x},t)
 \rho (\vec{y},s)} \eqno (9)$$
so that $\int \rho (\vec{y},s)p(\vec{y},s,\vec{x},t)d^3y=
\rho (\vec{x},t)$ 
and $\rho (\vec{y},s)=\int p_*(\vec{y},s,\vec{x},t)
\rho (\vec{x},t)d^3x$.  

The transport (density evolution) equations (6) and (7) refer to
processes running
in opposite  directions  in a fixed, common for both, time-duration period.
The  forward one, (6), executes an interpolation from the Borel set $A$
to $B$, while the  backward one, (7), executes  an interpolation from
$B$ to $A$, compare e.g. the defining identities (1).

The knowledge of the Feynman-Kac kernel (3) implies that the
transition probability density of the forward  process reads:
$${p(\vec{y},s,\vec{x},t)=k(\vec{y},s,\vec{x},t)
{{v(\vec{x},t)}\over {v(\vec{y},s)}}\enspace .}\eqno (10)$$
while the corresponding (derivable from (10), since
$\rho (\vec{x},t)$ is given) transition
probability density  of the backward process has the form:
$${p_*(\vec{y},s,\vec{x},t) = k(\vec{y},s,\vec{x},t)
{{u(\vec{y},s)}\over {u(\vec{x},t)}}\enspace .}\eqno (11)$$
Obviously, \cite{olk2,zambr}, in the time interval $0\leq s<t\leq T$
there holds:
$${u(\vec{x},t)=\int u_0(\vec{y}) k(\vec{y},s,\vec{x},t) d^3y}
\eqno (12)$$
$$v(\vec{y},s)=\int k(\vec{y},s,\vec{x},T) v_T(\vec{x})d^3x \enspace .
$$
Consequently, the system (4) fully determines the underlying
random motions, forward and backward, respectively.

\section{ The Burgers equation in Schr\"{o}dinger's interpolation}

The prototype nonlinear field equation named the Burgers or "nonlinear
diffusion" equation
(typically without, \cite{burg,hopf},
the forcing  term $\vec{F}(\vec{x},t)$):
$${\partial _t\vec{v}_B + (\vec{v}_B\nabla )\vec{v}_B = \nu
\triangle \vec{v}_B +
\vec{F}(\vec{x},t)}\eqno (13)$$
 recently  has acquired  a considerable popularity in the variety
 of physical contexts, \cite{prl}. \\
By dropping the force term in (13), we are left with a commonly used form
of the "nonlinear diffusion equation" whose solutions are known exactly, in
view of the Hopf-Cole linearising transformation mapping (13) into the
 heat equation. Here,
 $\partial _t\vec{v}_B + (\vec{v}_B\nabla )\vec{v}_B = \nu
\triangle \vec{v}_B$ is mapped into
$\partial _t \theta = \nu \triangle \theta $, by means of the substitution
$\vec{v}_B=-2\nu \nabla ln\theta $.  This linearisation of the Burgers
equation  is normally regarded to  be  devoid of any deeper
physical meaning,  and
specificaly the link with  stochastic processes determined by the
heat equation has not received a proper attention.
Our previous analysis shows that the intrinsic interplay between the
deterministic and random evolution, appropriate for a large class
of classically chaotic systems, extends to much wider framework.

Burgers velocity fields can be analysed on their own with different
(including random)   choices of the initial data and/or force fields.
However, we are interested in the possible diffusive matter transport
that is locally governed by Burgers flows, cf. \cite{prl}.
In this particular connection, let us point out a conspicouous
 hesitation that could have been  observed
in attempts to establish the  most appropriate  matter transport 
rule,  if any diffusion-type microscopic dynamics assumption is
 adopted to underly the "nonlinear diffusion" (13).\\
Depending  on the particular phenomenological departure point,
 one either adopts the standard continuity equation,
 \cite{zeld,alb}, that is certainly
valid to a high degree of  accuracy in the so-called low viscosity
limit $\nu \downarrow 0$ ,  but incorrect on
mathematical grounds \it if \rm  there is a  genuine Markovian
diffusion  process involved \it and \rm simultaneously a solution
of (13) stands
for the respective \it current \rm velocity of the flow: 
${\partial _t\rho (\vec{x},t)= - \nabla 
[\vec{v}(\vec{x},t)\rho (\vec{x},t)]\enspace . }
$ 

Alternatively, following  the white noise calculus tradition telling
that  the stochastic integral
$\vec{X}(t)=\int_{0}^{t} \vec{v}_B(\vec{X}(s),s)ds +
\int_{0}^{t} \vec{\eta }(s)ds$
necessarily implies the Fokker-Planck equation, one is tempted to adopt:
${\partial _t\rho (\vec{x},t) = \nu \triangle \rho (\vec{x},t) - 
\nabla [\vec{v}_B(\vec{x},t)\rho (\vec{x},t)]}$
which is clearly problematic in view of the classic Mc Kean's
discussion
of the  propagation of  chaos for the Burgers equation,
\cite{kean,cald,osa} and the derivation of the stochastic 
"Burgers process" in this  context: 
"the fun begins in trying to describe this Burgers motion as the 
path of a tagged molecule in an infinite bath of like molecules", 
\cite{kean}.

To put things on the solid ground, let us consider a Markovian
diffusion  process, which is characterised by the
transition probability density
(generally inhomogeneous in space and time law of random
displacements)
$p(\vec{y},s,\vec{x},t)\, ,\, 0\leq s<t\leq T$, and the probability 
density $\rho (\vec{x},t)$  of its random variable 
$\vec{X}(t)\, ,\,  0\leq t \leq T$. The process is completely 
determined  by these data. For clarity of discussion, we do not
impose  any spatial boundary  restrictions, nor fix any concrete 
limiting value of $T$ which, in principle, can be moved to infinity.

Let us confine attention to processes defined by the standard
backward diffusion equation.
Under suitable restrictions (boundedness of involved
functions, their continuous differentiability) the function:
$${g(\vec{x},s)=
\int  p(\vec{x},s,\vec{y},T)g(\vec{y},T) d^3y }\eqno (14)$$
satisfies the  equation
$${- \partial _sg(\vec{x},s) = \nu \triangle g(\vec{x},s)  +
[\vec{b}(\vec{x},s)\nabla ]g(\vec{x},s)
\enspace .}\eqno (15)$$
Let us point out that the validity of (14) is known to be a \it
necessary
 \rm condition for the existence of a Markov diffusion process, whose
probability density $\rho (\vec{x},t)$ is to obey the Fokker-Planck 
equation (the forward drift $\vec{b}(\vec{x},t)$ replaces the
previously  utilized Burgers velocity  $\vec{v}_B(\vec{x},t)$).

The case  of particular interest, in the traditional 
nonequilibrium statistical physics literature,  
appears when $p(\vec{y},s,\vec{x},t)$
is  a \it fundamental solution \rm of (15) with respect to variables
$\vec{y},s$,  \cite{krzyz,fried,horst},
see however \cite{olk2} for  an analysis of alternative situations.
Then, the transition probability density satisfies \it also \rm
the second Kolmogorov (e.g. the Fokker-Planck) equation in the
remaining $\vec{x}, t$ pair of variables.  Let us emphasize that
these two equations form an adjoint pair of partial differential
equations, referring to the  slightly counterintuitive for
physicists, though transparent for
mathematicians, \cite{haus,fol,has,nel,zambr}, issue of time
reversal of diffusions.

We can consistently introduce the random variable of the process in the
form  $\vec{X}(t)=
\int_0^t\, \vec{b}(\vec{X}(s),s)\, ds + \sqrt {2\nu } \vec{W}(t) $.
Then, in view of the standard rules of the It\^{o} stochastic calculus,
\cite{nel1,nel,zambr},  we realise that for any smooth function
$f(\vec{x},t)$ of the random variable $\vec{X}(t)$ the 
conditional expectation value:
$$lim_{\triangle t\downarrow 0} {1\over {\triangle t}}\bigl [\int
p(\vec{x},t,\vec{y},t+
\triangle
t)f(\vec{y},t+\triangle t)d^3y - f(\vec{x},t)\bigr ] = 
(D_+f)(\vec{X}(t),t)=    \eqno (16) $$
$$=  (\partial _t + \vec{b}\nabla + \nu \triangle )f(\vec{x},t)
\enspace , $$
where $\vec{X}(t)=\vec{x}$,
determines  the forward drift $\vec{b}(\vec{x},t)$ of the process
(if we set components of $\vec{X}$ instead of $f$) and, moreover,
 allows to introduce the local field of (forward) accelerations
 associated with the
diffusion process, which we \it constrain \rm by demanding
(see e.g. Refs.
\cite{nel,nel1,zambr} for prototypes of such dynamical
constraints):
$${(D^2_+\vec{X})(t) =(D_+\vec{b})(\vec{X}(t),t) =(\partial _t
\vec{b} +
(\vec{b}\nabla )\vec{b} + \nu \triangle
\vec{b})(\vec{x},t)= \vec{F}(\vec{x},t)}\eqno (17)$$
where $\vec{X}(t)=\vec{x}$ and, at the moment arbitrary, 
function $\vec{F}(\vec{x},t)$ may be  interpreted as
an external forcing applied to  the diffusing system,
\cite{blanch}. 

By invoking (9), we can also  define the backward derivative of
the process in the
conditional mean (cf. \cite{blanch,vig,olk} for a discussion of
these concepts in case of the most traditional Brownian motion and
Smoluchowski-type diffusion processes)
$$lim_{\triangle t\downarrow 0} \, {1\over {\triangle t}}\bigl
[ \vec{x} - \int p_* (\vec{y},t-\triangle t,\vec{x},t)\vec{y} d^3y
\bigr ]= (D_-\vec{X})(t)=
\vec{b}_*(\vec{X}(t),t)
\eqno (18) $$
$$(D_-f)(\vec{X}(t),t) = (\partial _t + \vec{b}_* \nabla - \nu
\triangle )f(\vec{X}(t),t)$$
Accordingly, the backward version  of the acceleration field reads
$${(D^2_-\vec{X})(t) = (D^2_+\vec{X})(t) = \vec{F}(\vec{X}(t),t) }
\eqno (19) $$
where  in view of $\vec{b}_*= \vec{b} - 2\nu \nabla ln \rho $
we have  explicitly fulfilled  the \it forced Burgers equation  \rm  :
$${\partial _t\vec{b}_* +  (\vec{b}_*\nabla )\vec{b}_* - 
\nu \triangle \vec{b}_* 
= \vec{F}} \eqno (20)$$
and, \cite{nel,zambr,blanch},  under the gradient-drift field
assumption, $curl \, \vec{b}_*=0$, we deal
with $\vec{F}(\vec{x},t)=2\nu \nabla c(\vec{x},t)$ where the Feynman-Kac
potential (3) is explicitly involved.

Let us notice that the familiar (linearization of the nonlinear problem)
Hopf-Cole transformation,
\cite{hopf,flem}, of the Burgers equation into the 
generalised diffusion equation (yielding explicit solutions in the 
unforced case) has been explicitly used before (the second formula (4))
 in the framework of the
 Schr\"{o}dinger interpolation problem.
In fact, by defining $\Phi _*=log\, u$,
we immediately recover the traditional form of the Hopf-Cole
transformation 
 for Burgers velocity fields: $\vec{b}_*=-2\nu \nabla \Phi _*$.
In the standard considerations that allows to map a nonlinear
(unforced Burgers) equation  into a linear, heat, equation.
In the special case of the standard free Brownian motion, there 
holds $\vec{b}(\vec{x},t)=0$ while $\vec{b}_*(\vec{x},t)=-2\nu 
\nabla log\, \rho (\vec{x},t)$. 

Let us point out that the equation (7) is in fact the only
transport equation where the Burgers
velocity field  is allowed to be undisputably present, under the
diffusive scenario assumption, \cite{prl}.   The standard continuity
equation  is certainly inappropriate for nonzero values od the
diffusion  constant $\nu $.

\section{Reconstruction of the microscopic dynamics from the
probability density data: obstacles exemplified}

We have mentioned before, that
another version of the Schr\"{o}dinger boundary data problem, \cite{mis},
departs directly from a given
(Fokker-Planck-type)  probability density evolution and investigates
 the circumstances allowing to deduce a unique random process from
 this dynamics.  Surely, solutions of the Fokker-Plack equation  itself
 do not
 yet determine the    underlying stochastic process. Additional
 assumptions are always  necessary and a number of traps to be avoided.

 As a particular guide to those obstacles, we shall refer to  the
 familiar   free quantum evolution that is regarded as the time
 adjoint parabolic problem, exactly in the spirit of our previous
 discussion.

In our  previous paper,  \cite{olk}, the major conclusion was that in
order to give 
a definitive probabilistic description of the quantum dynamics as a 
\it unique \rm diffusion process solving Schr\"{o}dinger's
interpolation problem, a suitable Feynman-Kac semigroup must be
singled out. Let us
point out  that the measure preserving dynamics, permitted in the 
presence of conservative force fields, was investigated in
\cite{blanch}, see also \cite{carm,freid}.

The present analysis 
was performed quite generally and extends to the dynamics affected by 
time dependent external potentials, with no 
clear-cut discrimination between the nonequilibrium statistical
physics  and essentially quantum  evolutions.
The formalism of Section 1 encompasses both groups of problems.
Nevertheless, it is quite illuminating to see directly how sensitive,
even  in simplest cases, the formalism is with respect to any
attempt of
relaxing our previous assumptions and the Schr\"{o}dinger
interpolation  problem rules-of-the-game.
Specifically in the quantum domain,
where the seemingly trivial case of the free  evolution, which is 
nonstationary,  needs the general parabolic system (4)
 to be considered. Even worse, then the system (4)
displays a  nontrivial nonlinearity:
the parabolic equations are coupled by the
effective, solution dependent potential. At the first glance,
this feature might seem to 
exclude the existence of any conceivable Feynman-Kac 
(dynamical semigroup) kernel, and in consequence  any common-sense 
law of random displacements (i.e. the transition probability density) 
 governing the pertinent stochastic evolution.
Certainly, the existence of fundamental solutions in this case is
far from being obvious.

At this point, let us emphasize that our principal goal is to take  
seriously the Schr\"{o}dinger picture quantum dynamics under the 
premises of the Born statistical postulate.
Hence, once we select \it 
as appropriate \rm a concrete  quantal interpolation between the 
prescribed 
(phenomenologically supported in particular) input-output statistics
data  $\rho _0(x)$ and $\rho _T(x)$ in terms of $\rho (x,t)=
{\overline {\psi }}(x,t)\psi (x,t),\; t\in [0,T]$, where $\psi (x,t)$
solves the Schr\"{o}dinger equation then,
on exactly the same footing, we are entitled to  look for 
an alternative probabilistic  explanation (or \it appropriate \rm  
description) of the very same interpolation, in terms of a well
defined  Markov stochastic (eventually diffusion)
process.

We shall proceed in the spirit of Section 1, while restricting our
discussion to the free Schr\"{o}dinger dynamics. Following
Ref. \cite{olk} we shall discuss the rescaled problem so as to 
eliminate all dimensional constants.

The free Schr\"{o}dinger evolution $i\partial _t\psi =
-\triangle \psi $ 
implies  the following propagation of a specific Gaussian
wave packet:
$${\psi (x,0)=(2\pi )^{-1/4} exp\: (-{{x^2}\over {4}})\; 
\longrightarrow  }\eqno (21) $$
$$\psi (x,t)=({2\over \pi })^{1/4} \; (2+ 2it)^{-1/2} 
exp[-{x^2\over {4(1+it)}}]$$
So that 
$${\rho _0(x)=|\psi (x,0)|^2=(2\pi )^{-1/2}\: exp[-{x^2\over 2}] 
\longrightarrow }\eqno (22)$$
$$\rho (x,t)=|\psi (x,t)|^2= [2\pi (1+t^2)]^{-1/2}\: 
exp [-{x^2\over {2(1+t^2)}}]$$ 
and the Fokker-Planck equation (easily derivable from the standard 
continuity equation $\partial _t\rho =-\nabla (v\rho ),\; v(x,t)=
xt/(1+t^2)$) holds true:
$${\partial _t\rho = \triangle \rho - \nabla (b\rho )\; \;  , \; \; 
b(x,t)= - {{1-t}\over {1+t^2}}\: x} \eqno (23)$$

The Madelung factorization $\psi =exp(R+iS)$ implies (notice that 
$v=2\nabla S$ and $b=2\nabla (R+S)$) that the 
related real functions  $\theta(x,t)=exp(R+S)$ and $\theta _*(x,t)=
exp(R-S)$ read:
$$\theta (x,t)=[2\pi (1+t^2)]^{-1/4} exp(-{x^2\over 4}\: 
{{1-t}\over {1+t^2}} - {1\over 2} arctan\: t)$$
$${\theta _*(x,t)=[2\pi (1+t^2)]^{-1/4} exp(-{x^2\over 4}\: 
{{1+t}\over {1+t^2}} + {1\over 2} arctan\: t)}\eqno (24)$$
They solve a suitable version  of the general parabolic equations (4),
namely :
$${\partial _t \theta =-\triangle \theta + {1\over 2} \Omega  \theta }
\eqno (25)$$
$$\partial _t\theta _* =\triangle \theta _* - {1\over 2} 
\Omega \theta  _*$$
with 
$${{1\over 2}\Omega (x,t) = {x^2\over {2(1+t^2)^2}} -
{1\over {1+t^2}} = 
2{{\triangle \rho ^{1/2}}\over {\rho ^{1/2}}}= Q(x,t)}\eqno (26)$$

By setting $t=T$ we associate with the above dynamics  the terminal 
density $\rho _T(x)$, and then the concrete 
Schr\"{o}dinger boundary data problem for the stochastic
interpolation $\rho _0(x)\rightarrow \rho _T(x)$, (1).

To capture the spirit of our previous  discussion, we shall
replace  equations (25) by the more general equations (4), where
only the potential $c(x,t)$ will be identified with the above
${1\over 2}\Omega (x,t)$. Then, we shall look for solutions
$u(x,t),\: v(x,t)$ of
these parabolic equations, and  in particular we shall identify
the quantally implemented functions $\theta _*(x,t),
\theta (x,t)$, (24), \it among \rm them.
Effectively, it amounts to the previously mentioned
linearisation of the nonlinear parabolic system.

In view of the relatively simple form of the  probability density  
$\rho (x,t)$, (22)  one might be tempted to guess (more or less
fortunately)  the transition probability
density, consistent with the propagation (22). However,
it is well known
that there are many stochastic    processes implying (22) for all
$t\in [0,T]$, which not necessarily have 
much in common  with the original wave function 
dynamics (21).
In general they are  incompatible  with the corresponding
parabolic system (cf. (4) and (25)).
If it happens otherwise, the reason for this proliferation of
would-be consistent stochastic processes is rooted in exploting the
particular functional form of solutions, instead of relying on the 
form-independent arguments, e.g. (4).

Let us consider simple examples which, albeit coming under very 
special circumstances (free dynamics with a specific initial wave
packet  choice, and no zeros admitted in the course of the
propagation), clearly idicate how important is the \it proper
choice  \rm of the Feynman-Kac kernel.
The virtue of a parabolic system (4) is that its form
is universal for the Schr\"{o}dinger dynamics, and thus does not
depend on  a particular functional form of solutions nor this of
external potentials.
It appears that the system (4) sets a very rigid
framework for the  probabilistic manifestations (e.g. stochastic
processes)  of the quantum Schr\"{o}dinger dynamics. \\

{\bf Example 1}: We shall demonstrate that an improper (\it not \rm 
through (4) or (25)), but
fortunate, choice of the kernel might lead to an alternative
stochastic  representation of the quantum dynamics (22).\\
Let us begin from directly introducing the transition probability
density
$${p(y,s,x,t)=[2\pi (t^2-s^2)]^{-1/2}\; exp [-{{(x-y)^2}
\over {2(t^2-s^2)}}]}
\eqno (27)$$
which for all intermediate  times $0\leq s<t\leq T$ executes
a desired 
propagation $\rho (x,t)=\int p(y,s,x,t)\rho (y,s)dy$, (22).
Clearly, the Chapman-Kolmogorov identity
$\int p(y,s,z,\tau )p(z,\tau ,x,t)
d\tau =p(y,s,x,t)$ holds true, and the properties (the first one,
for all  $\epsilon >0$):
$$\lim_{\triangle t\downarrow 0} {1\over {\triangle t}} 
\int_{|x-y|>\epsilon }\: p(y,t,x,t+\triangle t) dx = 0$$
$${lim_{\triangle t\downarrow 0}{1\over {\triangle t}}
\int_{-\infty }^{+\infty } (x-y)\: 
p(y,t,x,t+\triangle t)dx = 0 }\eqno (28)$$
$$lim_{\triangle t\downarrow 0}{1\over {\triangle t}}
\int_{-\infty }^{+\infty } (x-y)^2\: 
p(y,t,x,t+\triangle t)dx= 2t$$ 
tell us that the law of random displacements $p(y,s,x,t)$, (27), can
be attributed to a Markov diffusion process associated with the
parabolic (Fokker-Planck) equation
$${\partial _t \rho = t\triangle _x \rho }\eqno (29)$$
In fact, our $p(y,s,x,t)$ is a fundamental solution of this equation
 with  respect to $x, t$ variables,
 while obeying the  time adjoint parabolic
equation in the remaining (e.g. $y, s$) pair of variables
$${\partial _sp(y,s,x,t)=-s\triangle _yp(y,s,x,t)}\eqno (30)$$
This diffusion has  a vanishing forward drift and the quadratic in
time  variance (the diffusion coefficient equals $t$),
hence its local
characteristics  are completely divorced from those of the 
Nelson process \cite{olk} derivable from the solution (21) of the
Schr\"{o}dinger equation.

Interestingly, since $p(y,s,x,t)$ itself is a perfect,
strictly positive and 
continuous in all variables (Markov) semigroup kernel,
nothing prevents  
us from performing the Schr\"{o}dinger problem analysis (1) with  the
boundary densities $\rho _0(x)$ and $\rho _T(x)$ defined by the above
free evolution problem.
However, we shall proceed otherwise and having given 
explicit solutions of the parabolic system (25) we introduce another
strictly positive and continuous in all variables  function:
$${k_1(y,s,x,t)=p(y,s,x,t){{\theta (y,s)}\over {\theta (x,t)}}=
[2\pi (t^2-s^2)]^{-1/2} ({{1+t^2}\over {1+s^2}})^{1/4} }\eqno (31)$$
$$exp[-{{(x-y)^2}\over {2(t^2-s^2)}}]\: 
exp[-{y^2\over 4}{{1-s}\over {1+s^2}} + {x^2\over 4}{{1-t}
\over {1+t^2}}]\; exp[{1\over 2}(arctan\: t-arctan\: s)]$$
and  observe that the Schr\"{o}dinger system (1) in the present
situation  is involved as well, since trivially there holds:
$${\rho _0(x)=\theta _*(x,0)\int k_1(x,0,y,T)\theta (y,T)dy}
\eqno (32)$$
$$\rho _T(x)=\theta (x,T)\int k_1(y,0,x,T)\theta _*(y,0)dy$$

Disregarding the derivation which has led us to (22), we can simply
consider (22) as the Schr\"{o}dinger system of equations with a
fixed kernel
and boundary density data. Then, we immediately infer that
by Jamison's theorem, \cite{jam},
its unique (up to a coefficient) solution 
is constituted by the pair $\theta _*(x,0),\: \theta (x,T)$
of functions,  already determined by (24).
Moreover, $k_1(y,s,x,t)$ obeys the Chapman-Kolmogorov composition
rule:
$${\int k_1(y,s,z,\tau )k_1(z,\tau ,x,.t)d\tau =}\eqno (33)$$
$$\int p(y,s,z,\tau ){{\theta (y,s)}\over {\theta (z,\tau )}}
p(z,\tau ,x,t) 
{{\theta (z,u)}\over {\theta (x,t)}}\: d\tau =
p(y,s,x,t) {{\theta (y,s)}\over {\theta (x,t)}}=k_1(y,s,x,t)$$
In view of $\int p(y,s,x,t)dx=1$ for all $s<t$, we have
$${\int k_1(x,s,y,t)\theta (y,t)dy = \theta (x,s)}\eqno (34)$$
and, since $\theta \theta _*=\rho $, we get 
$${\int k_1(y,s,x,t)\theta _*(y,s)dy=\int \theta _*(y,s)p(y,s,x,t)
{{\theta (y,s)}\over {\theta (x,t)}}dy=}\eqno (35)$$
$${1\over {\theta (x,t)}}\int \rho (y,s)p(y,s,x,t)dy=
{{\rho (x,t)}\over {\theta (x,t)}}=\theta _*(x,t)$$
Thus, undoubtedly we have in hands a complete solution of 
the Schr\"{o}dinger boundary data problem (1):
for the once chosen  kernel $k_1$, this solution
is  unique, and compatible  with the dynamics of the
corresponding  Schr\"{o}dinger wave function.
But, the constructed stochastic process
is completely incongreuent with the standard wisdom about
Nelson's diffusion  processes \cite{zambr,nel,nel1,olk}.
The reason is clear: our analysis was performed  for a
particular solution, whose functional form allows for an alternative 
stochastic representation. But, let us stress the point, if we look
for the functional-form-independent construction, it is the
parabolic system (4) from which one should depart.

Anyway, even  the inappropriate choice of the integral kernel $k_1$,
does  allow  to derive the quantum mechanically implemented
dynamics (22) from
respectively $\theta (x,T)$ and $\theta _*(x,0)$, by means of the 
propagation formulas (4). The probability density evolves in time
correctly, but the vanishing drift and the linear in time 
diffusion coefficient situate this stochastic process outside 
the scope set by (25)  and (4).\\

{\bf Example 2}: We shall demonstrate, that another choice of the
kernel, still with no reference to the system (4),
will allow to reproduce the
stochastic propagation with the probability density,
drifts  and diffusion 
coefficient of Nelson's stochastic mechanics, which  however is 
\it not \rm Nelson's process for the quantum evolution (22).\\
We are inspired by our previous paper \cite{olk},
where an interesting 
stochastic propagation, compatible with (22), was introduced
by means of the transition probability density:
$$p_{y,s}(x,t)=[4\pi (t-s)]^{-1/2}\;
exp[-{{(x-c_{t,s}y)^2}\over {4(t-s)}}]$$
$${p_{y,s}(x,s)=\delta (x-y)\; ,\; 0\leq s<t\leq T}\eqno (36)$$
$$c_{t,s}= [{{(1-t)^2+2s}\over {1+s^2}}]^{1/2}$$
Here, the  density $\rho (y,s)$, (22), is propagated into
the corresponding $\rho (x,t)$ according to the rule 
$\rho (x,t)=\int p_{y,s}(x,t)\rho (y,s)dy$, for all intermediate 
times $0\leq s<t\leq T$.
As noticed in \cite{olk}, this propagation is somewhat pathological
since 
it \it does not \rm obey the Chapman-Kolmogorov  composition rule: 
$\int p_{y,s}(z,\tau )p_{z,\tau }(x,t)d\tau \not= p_{y,s}(x,t)$ 
and thus $p$ cannot be interpreted as a transition density of the
Markov  process.

However, if we would naively proceed like  in the Example 1 and
define the strictly positive continuous function
$${k_2(y,s,x,t)=p_{y,s}(x,t){{\theta (y,s)}\over {\theta (x,t)}}}
\eqno (37)$$
where $0\leq s<t\leq T$ and $\theta (x,t)$ is given by (24), then the
Schr\"{o}dinger system (32), with $k_2$ replacing $k_1$,
trivially appears.
Indeed, because of $\int p_{y,s}(x,t)dx=1$  for $s<t$, there holds:
$${\theta _*(x,0)\int k_2(x,0,y,T)\theta (y,T)dy= 
\theta _*(x,0)\int p_{x,0}(y,T)\theta (x,0)dy=}$$
$$=\theta _*(x,0)\theta (x,0)=\rho _0(x)$$
$${\theta (x,T)\int k_2(y,0,x,T)\theta _*(y,0)dy = 
\int p_{y,0}(x,T)\theta (y,0)\theta _*(y,0)dy =}\eqno (38)$$
$$=\int p_{y,0}(x,T)\rho (y,0)dy=\rho_T(x) $$
As a  consequence, if we analyze the above Schr\"{o}dinger system
with the  boundary data $\rho _0(x)$ and $\rho _T(x)$  fixed by (22)
(as before),
but with the new  kernel $k_2$ then, somewhat unexpectedly, the same as
before pair $\theta (x,0),\theta _*(x,T)$
necessarily comes out as a solution.
Let us emphasize that the solution is 
unique for the chosen kernel $k_2$, albeit it coincides  with the 
unique (as well) 
solution previously associated with the kernel $k_1$ (cf. Example 1).

The meaning of the uniqueness of solution of the Schr\"{o}dinger system 
\cite{jam} becomes clear:
if we have prescribed the boundary density data 
the solution is unique for a chosen kernel, 
but there are many kernels which may  give rise to the very same
solution.

The pathology (non-Markovian density) of $p_{y,s}(x,t)$ extends to 
$k_2(y,s,x,t)$ and the semigroup composition rule is invalid in 
this case. Nevertheless, we can blindly  repeat the step (32),
 with $k_2$ instead of $k_1$, so reproducing the evolution (22).
Moreover, in the present case, \cite{olk},
we can exploit the standard 
recipe  to evaluate the forward drift of a conventional diffusion:
$${lim_{\triangle t\downarrow 0}{1\over {\triangle t}} 
[\int y\: p_{x,t}(y,t+\triangle t)dy \: -\: x]= b(x,t)=
-{{1-t}\over {1+t^2}}\: x}\eqno (39)$$
Clearly, it is the forward drift of the Nelson diffusion
\cite{nel,olk}  associated with (24), and it  consistently
appears in the corresponding Fokker-Planck equation (6).\\

Let us observe that $p_{y,s}(x,t)$  solves  the first
Kolmogorov equation with respect to $x,t$:
$${\partial _tp_{y,s}(x,t)=\triangle _xp_{y,s}(x,t) - b_{y,s}(t)
\nabla _xp_{y,s}(x,t)}\eqno (40)$$
$$b_{y,s}(t)=y{{\partial c_{t,s}}\over {\partial t}}$$
As such, it can be exploited to construct a genuine Markov process,
albeit  disconnected from the quantal dynamics (22). Namely, we
can define another solution of the  equation (40), in variables
$x_1,t_1$:
$${p_{y,s} (x_1,t_1,x_2,t_2)=[4\pi (t_2-t_1)]^{-1/2} 
exp[-{{(x_2-x_1-c\: y)^2}\over {4(t_2-t_1)}}]}\eqno (41)$$
$$c=c_{t_2,s} - c_{t_1,s}\; , \; 0\leq s<t_1 < t_2\leq T$$
with $c_{t,s}$ given  by (36).
It is easy to verify that the transition density (41) actually
\it is \rm a fundamental solution, and as such satisfies the second  
Kolmogorov equation with respect to $x_2,t_2$, for each fixed  
$y,s$ label, $0\leq s<t_1<t_2\leq T$.
Consequently, we have in hands the $(y,s)$-family of well defined 
Markovian transition probability densities $p_{y,s}$ for random
propagation  scenarios.
Indeed, to this end one needs to check the  (apparent)  
compatibility conditions:
(a) $p_{y,s}(x_1,t,x_2,t)=\delta (x_2-x_1)$, \\
(b) $\int p_{y,s}(x_1,t_1,x_2,t_2)p_{y,s}(x_1,t_1)dx_1 =
p_{y,s}(x_2,t_2)$  and  in addition  \\
(c) $\int p_{y,s}(x_1,t_1,x_2,t_2)p_{y,s}(x_2,t_2,x_3,t_3)dx_2 = 
p_{y,s}(x_1,t_1,x_3,t_3)$,
where $0\leq s<t_1<t_2<t_3\leq T$ and $p_{y,s}(x,t) $ plays the
role of the  density of the Markov process.
 The identity (c) in the above is the Chapman-Kolmogorov formula.
  
To avoid the above obstacles, the only how-to-proceed procedure
 is provided by the route outlined before, e.g.  that leading
from the Feynman-Kac kernel to the associated Markov diffusion
process via Schr\"{o}dinger's boundary-data problem. 
The complete solution to this particular issue, in the quantum
dynamics context, has been given elswhere, \cite{olk}.

\vskip0.2cm
{\bf Acknowledgement}: P. G. and R. O.  received a financial support from
the  KBN research grant No 2 P302 057 07.

\end{document}